\documentclass[preprint,aps,eqsecnum,superscriptaddress]{revtex4}
\usepackage[T1]{fontenc}
\usepackage[latin1]{inputenc}
\usepackage{graphicx,amsfonts}
\usepackage{amsmath}
\usepackage{slashed}
\newcommand{\be}{\begin{equation}}
\newcommand{\ee}{\end{equation}}
\newcommand{\bea}{\begin{eqnarray}}
\newcommand{\eea}{\end{eqnarray}}
\newcommand{\bml}{\begin{subequations}}
\newcommand{\eml}{\end{subequations}}

\newcommand{\e}{\epsilon}

\newcommand{\Ket}[1]{\left|\, #1 \,\right\rangle }

\newcommand{\rmi}{\mathrm{i}}

\begin{document}

\title {Noncommutative quantum mechanics as a gauge theory}

\author{F. S. Bemfica}
\author{H. O. Girotti}
\affiliation{Instituto de F\'{\i}sica, Universidade Federal do Rio Grande
do Sul, Caixa Postal 15051, 91501-970 - Porto Alegre, RS, Brazil}
\email{fbemfica, hgirotti@if.ufrgs.br}

\begin{abstract}
The classical counterpart of noncommutative quantum mechanics is a constrained system containing only second class constraints. The embedding procedure formulated by Batalin, Fradkin and Tyutin (BFT) enables one to transform this system into an Abelian gauge theory exhibiting only first class constraints. The appropriateness of the BFT embedding, as implemented in this work, is verified by showing that there exists a one to one mapping linking the second class model with the gauge invariant sector of the gauge theory. As is known, the functional quantization of a gauge theory calls for the elimination of its gauge freedom. Then, we have at our disposal an infinite set of alternative descriptions for noncommutative quantum mechanics, one for each gauge. We study the relevant features of this infinite set of correspondences. The functional quantization of the gauge theory is explicitly performed for two gauges and the results compared with that corresponding to the second class system. Within the operator framework the gauge theory is quantized by using Dirac's method.
\end{abstract}

\maketitle
\newpage

\section{Introduction}
\label{sec:level1}

We shall be concerned here with quantum systems whose dynamics is described by a self-adjoint Hamiltonian operator ${\hat H}({\hat Q},{\hat P})$ made up of the Cartesian coordinate operators ${\hat Q}^j, j= 1,\ldots, N$ and their canonically conjugate momenta ${\hat P}_j, j = 1,\ldots, N$. However, unlike the usual case, coordinates and momenta operators are supposed to obey the non-canonical equal-time commutation rules

\bml
\label{I-1}
\bea
&&\left[{\hat Q}^l, {\hat Q}^j\right] = -2 \rmi\hbar\theta^{lj}\,,\label{mlett:aI-1}\\
&&\left[{\hat Q}^l, {\hat P}_j\right] = \rmi\,\hbar\,\delta^{l}_{\,\,j}\,,\label{mlett:bI-1}\\
&&\left[{\hat P}_l, {\hat P}_j\right] = 0\,.\label{mlett:c1}
\eea
\eml

\noindent
The distinctive feature is that the coordinate operators do not commute among themselves. The lack of non-commutativity of the coordinates is parameterized by the real antisymmetric $N \times N$ constant matrix $\|\theta\|$. In Refs.\cite{Chaichian1,Gamboa1,Gamboa2,Girotti1,Bemfica} one finds specific examples of noncommutative systems whose quantization has been carried out. The conditions for the existence of the Born series and unitarity were investigated in Ref.\cite{Bemfica1} while a general overview of the connection linking noncommutative theories with constrained systems was presented in \cite{Bemfica2}. As for the uniqueness of the functional description, when using the time-slice definition for the phase space path integral, it was established in Ref.\cite{Bemfica3}.

The classical counterpart of a quantum system involving non-commuting coordinates must necessarily be a constrained system \cite{footnote2}. Indeed, the equal time algebra in Eq.(\ref{mlett:aI-1}) could not have been abstracted from a Poisson bracket algebra, simply because the Poisson bracket of two coordinates vanishes. Now, the problem of finding a constrained system mapping onto the noncommutative theory specified in (\ref{I-1}) has already been solved \cite{Deriglazov1}. Its classical dynamics is described by the Lagrangian \cite{footnote3}

\be
\label{I-2}
 L\,=\,a\,v_j\,{\dot q}^j\,-\,h_0(q^j , a v_j)\,+\,a^2\,{\dot v}^i\,\theta_{ij}\,v^j\,,
\ee

\noindent
where repeated Latin indices sum from 1 to $N$. The constraint structure of this system reduces to the primary second-class constraints

\bml
\label{I-3}
\bea
&& G_i\,\equiv\,p_i\,-\,a\,v_i\,\approx\, 0\,,\label{mlett:aI-3}\\
&& T_i\,\equiv\,\pi_i\,-\,a^2\,\theta_{ij} v^j\,\approx\,0\,,\label{mlett:bI-3}
\eea
\eml

\noindent
where $p_i$ ($\pi_i$) is the momentum canonically conjugate to the generalized coordinate $q^i$ ($v^i$) and the sign of weak equality ($\approx$) is being used in the sense of Dirac \cite{Dirac1}. As for the canonical Hamiltonian one finds that

\be
\label{I-4}
H^{(0)}(q , p)\,=\,h_0(q , p)\,.
\ee

\noindent
We shall substantiate in the next section the fact that the equal-time algebra in Eq.(\ref{I-1}) is, in fact, the quantum counterpart of the Dirac bracket algebra arising in connection with the model specified in Eq.(\ref{I-2}).

Now, it is known that by appropriately enlarging the phase space of a second class theory one obtains a first class one (a gauge theory). Then, the formulation of the resulting gauge theory in different gauges yields different realizations of the initial second class model. In Section II we display and discuss the results arising from applying the embedding procedure of Batalin, Fradkin and Tyutin (BFT) \cite{Bat,Frad,Tyu} to the second class system that gives origin to noncommutative quantum mechanics, namely, the one specified by the Lagrangian in Eq.(\ref{I-2}). The result is an Abelian gauge theory whose gauge invariant sector maps isomorphically onto the second class model. This is shown in Section III. Section IV is dedicated to formulate the phase space functional quantization of the second and first class theories. The gauge theory is quantized in different gauges and the results compared with those obtained for the second class one. In section V we use the method put forward by Dirac \cite{Dirac1} to implement the operator quantization of the BFT extension of noncommutative quantum mechanics. The outcomes from this procedure will be shown to be in agreement with the results obtained from the operator quantization of the initial second class model. Section VI contains the conclusions.

\section{BFT embedding}
\label{sec:level2}

Our estrategy follows closely that in Refs.\cite{Bat,Frad,Tyu,Fleck}. We start by compactifying the notation as follows

\bea
\label{II-1}
q^{\mu}\,\equiv\,\left\{
\begin{array}{l}
q^j,\,1\leq \mu \leq N \\
v^j,\,N + 1 \leq\mu \leq 2N
\end{array}
\right.\,,
\eea

\bea
\label{II-2}
p_{\mu}\,\equiv\,\left\{
\begin{array}{l}
p_j,\,1\leq \mu \leq N \\
\pi_j,\,N + 1 \leq\mu \leq 2N
\end{array}
\right.\,.
\eea

\noindent
Clearly, $p_{\mu}$ is the momentum canonically conjugate to the coordinate $q^{\mu}$ obeying the canonical Poisson bracket ($PB$) algebra $[q^{\mu}\,,\,p_{\nu}]_{PB} = \delta^{\mu}_{\,\,\nu}$.

We introduce, in the sequel, the singular $2N \times 2N$ matrix

\be
\label{II-3}
||\Theta||\equiv
\left[\begin{array}{c|c}
0_N & a g_{\mu\,,\,\nu-N} \\
\hline
0_N & a^2\theta_{\mu-N\,,\nu-N}
\end{array}
\right]\,,
\ee

\noindent
where repeated Greek indices sum from $1$ to $2N$ and  $g_{\mu \nu}$ designates the Euclidean metric of the $2N$ dimensional space. One can verify that

\be
\label{II-4}
{\mathcal T}^{(0)}_\mu \equiv p_\mu-\Theta_{\mu\nu} q^\nu =\begin{cases}
G_i, \; 1\le\mu\le N\\
T_i, \; N+1\le\mu\le 2N
\end{cases}\,.
\ee

\noindent
Hence, the elements ($\Delta_{\mu \nu}$) of the antisymmetric non-singular Faddeev-Popov matrix ($\|\Delta\|$) are found to read

\be
\label{II-5}
\Delta_{\mu\nu}\,\equiv\,[{\mathcal T}^{(0)}_\mu\,,{\mathcal T}^{(0)}_\nu]_{PB}\,=\,\Theta_{\nu\mu}-\Theta_{\mu\nu}\,=\,-\Delta_{\nu\mu}
\ee

\noindent
or, more explicitly,

\be
\label{II-6}
\|\Delta\|\,=\,\left[\begin{array}{c|c}
0_N & -ag_{\mu\,,\,\nu-N}\\
\hline
ag_{\mu-N\,,\,\nu} & -2a^2\theta_{\mu-N\,,\,\nu-N}
\end{array}\right]\,.
\ee

We have already at hand all the ingredients needed for computing the basic Dirac brackets ($DB$). We skip the details and quote

\be
\label{II-7}
\left[q^\mu,q^\nu\right]_{DB}=(\Delta^{-1})^{\mu\nu}\Longrightarrow
\begin{cases}
\left[q^i,q^j\right]_{DB}=-2\theta^{ij}\\
\left[q^i,v^j\right]_{DB}=\frac{1}{a}g^{ij}\\
\left[v^i,v^j\right]_{DB}=0
\end{cases}\,,
\ee

\be
\label{II-8}
\left[q^\mu,p_\nu\right]_{DB}=\delta^\mu_\nu+(\Delta^{-1})^{\mu\alpha}\Theta_{\alpha\nu}\Longrightarrow
\begin{cases}
\left[q^i,p_j\right]_{DB}=\delta^i_j\\
\left[q^i,\pi^j\right]_{DB}=-a\theta^{ij}\\
\left[v^i,p_j\right]_{DB}=0\\
\left[v^i,\pi_j\right]_{DB}=0
\end{cases}\,,
\ee

\be
\label{II-9}
\left[p_\mu,p_\nu\right]_{DB}=\Theta_{\alpha\mu}(\Delta^{-1})^{\alpha\beta}\Theta_{\beta\nu}=0 \Longrightarrow
\begin{cases}
\left[p_i,p_j\right]_{DB}=0\\
\left[p_i,\pi_j\right]_{DB}=0\\
\left[\pi_i,\pi_j\right]_{DB}=0
\end{cases}\,,
\ee

\noindent
where

\be
\label{II-91}
\|\Delta\|^{-1} \,=\,\left[\begin{array}{c|c}
-2\theta_{\mu\,,\,\nu} & \frac{1}{a}g_{\mu\,,\,\nu-N}\\
\hline
-\frac{1}{a}g_{\mu-N\,,\,\nu} & 0_N
\end{array}\right]\,.
\ee

Within the $DB$ algebra the constraints hold as strong identities and may be used, for instance, to eliminate from the game the sector of the phase space spanned by the variables $v$ and $\pi$. As for any two functions of the remaining variables, $f(q , p)$ and $g(q , p)$ say, the correspondence rule

\be
\label{II-10}
[{\hat F}\,,\,{\hat G}]\,=\,i\,\hbar\,[f\,,\,g]_{DB}\bigg|_{\begin{array}{l} q \rightarrow {\hat Q}\\ p  \rightarrow {\hat P}\end{array}}
\ee

\noindent
provides us with a faithful quantization procedure. As usual, a supplementary ordering prescription may be needed. We emphasize that this rule, together with Eqs.(\ref{II-7})-(\ref{II-9}), allows for recovering the equal-time commutator algebra in Eq.(\ref{I-1}) and, therefore, confirms the assertion made in Section \ref{sec:level1} about the Lagrangian in Eq.(\ref{I-2}) being the classical counterpart of noncommutative quantum mechanics.

The first step towards the BFT embedding of the second class system under scrutiny consists in enlarging the original phase space by adding $2N$ new coordinates ($u^{\mu}$) and their corresponding canonically conjugate momenta ($s_{\mu}$) \cite{footnote4}. The quantities of interest are, nevertheless, the composite variables

\be
\label{II-11}
z^{\mu}\,\equiv\,-\,\frac{1}{2}\,u^{\mu}\,+\,\omega^{\mu \nu} s_{\nu}\,,
\ee

\noindent
where $\omega^{\mu \nu}$ is a $2N \times 2N$ real constant matrix which, so far, remains at our disposal. Since $[u^{\mu}\,,\,s_{\nu}]_{{PB}_{\Lambda}}\,=\,\delta^{\mu}_{\,\,\nu}$ one obtains \cite{footnote5}

\be
\label{II-12}
[z^{\mu}\,,\,z^{\nu}]_{{PB}_{\Lambda}}\,=\,(\Delta^{-1})^{\mu \nu}\,,
\ee

\noindent
where we have chosen, once and for all,

\be
\label{II-13}
\omega^{\mu \nu}\,\equiv\,(\Delta^{-1})^{\mu \nu}\,.
\ee

We shall seek next for extensions of the constraints, ${\mathcal T}_{\mu}^{(0)}(q,p) \rightarrow {\mathcal T}_{\mu}(q,p,z)$, and of the Hamiltonian, $H^{(0)}(q,p) = h_0(q,p) \rightarrow {\mathcal H}(q,p,z)$, verifying the strong involution algebra

\bml
\label{II-14}
\bea
&&[{\mathcal T}_{\mu}(q,p,z)\,,\,{\mathcal T}_{\nu}(q,p,z)]_{{PB}_{\Gamma}}\,=\,0\,,\label{mlett:aII-14}\\
&&[{\mathcal T}_{\mu}(q,p,z)\,,\,{\mathcal H}(q,p,z)]_{{PB}_{\Gamma}}\,=\,0\,,\label{mlett:bII-14}
\eea
\eml

\noindent
and the boundary conditions ${\mathcal T}_{\mu}(q,p,z = 0)\,=\,{\mathcal T}^{(0)}_{\mu}(q,p)$, ${\mathcal H}(q,p,z = 0)\,=\,H^{(0)}(q,p)$. By definition, Eqs.(\ref{II-14}) specify an Abelian gauge theory \cite{Barcelos}. In particular, we are interested in extensions of the form \cite{Bat,Frad,Tyu,Fleck}

\bml
\label{II-15}
\bea
&&{\mathcal T}_{\mu}(q,p,z)\,=\,{\mathcal T}^{(0)}_{\mu}(q,p)\,+\,\sum_{M = 1}^{+\infty} {\mathcal T}^{(M)}_{\mu}(q,p,z)\,,\label{mlett:aII-15}\\
&& {\mathcal H}(q,p,z)\,=\,H^{(0)}(q,p)\,+\,\sum_{M = 1}^{+\infty} H^{(M)}(q,p,z)\,,\label{mlett:bII-15}
\eea
\eml

\noindent
where

\be
\label{II-16}
{\mathcal T}^{(M)}_{\mu}(q,p,z)\,=\,X^{(M)}_{\mu \alpha_1...\alpha_M}(q , p)\,z^{\alpha_1}...z^{\alpha_M}\,,
\ee

\noindent
and

\be
\label{II-17}
H^{(M)}(q,p,z)\,=\,Y^{(M)}_{\alpha_1...\alpha_M}(q , p)\,z^{\alpha_1}...z^{\alpha_M}\,.
\ee

\noindent
The problem consists, of course, in determining $X$ and $Y$.

We concentrate first on Eq.(\ref{mlett:aII-14}). By substituting Eq.(\ref{II-16}) and Eq.(\ref{mlett:aII-15}) into Eq.(\ref{mlett:aII-14}) and, then, isolating the terms of order $z^{0}$ one obtains

\be
\label{II-18}
\Delta_{\mu \nu}\,+\,X^{(1)}_{\mu \alpha}\,(\Delta^{-1})^{\alpha \beta}\,X^{(1)}_{\nu \beta}\,=\,0\,,
\ee

\noindent
where we have taken into account Eq.(\ref{II-5}). It is clear that

\be
\label{II-19}
X^{(1)}_{\mu \nu}\,=\,\Delta_{\mu \nu}\,=\,-\,\Delta_{\nu \mu}
\ee

\noindent
solves Eq.(\ref{II-18}). The relevant point to be noticed in connection with this solution is that $X^{(1)}_{\mu \nu}$ does not depend on $q$ and/or $p$. This implies that

\be
\label{II-20}
[{\mathcal T}^{(0)}_{\mu}(q,p)\,,\,{\mathcal T}^{(1)}_{\nu}(q,p,z)]_{{PB}_{\Sigma}}\,=\,0\,,
\ee

\noindent
which in combination with the symmetry assumptions

\bml
\label{II-21}
\bea
&&X^{(M)}_{\mu \alpha_1 \ldots \alpha_j \ldots \alpha_k \ldots \alpha_M}(q , p) \,=\,+\,X^{(M)}_{\mu \alpha_1 \ldots \alpha_k \ldots \alpha_j \ldots \alpha_M}(q , p)\,,\label{mlett:aII-21}\\
&&X^{(M)}_{\mu \alpha_1 \ldots \alpha_j \ldots \alpha_k \ldots \alpha_M}(q , p)\,=\,-\,X^{(M)}_{\alpha_j  \alpha_1 \ldots \mu \ldots \alpha_k \ldots \alpha_M}(q , p)\,,\,\,\,\,\forall\,\,\,\alpha_j\,,\label{mlett:bII-21}
\eea
\eml

\noindent
yields

\be
\label{II-22}
X^{(M)}_{\mu \alpha_1 \ldots \alpha_j \ldots \alpha_k \ldots \alpha_M}(q , p) \,=\,0 \Longrightarrow {\mathcal T}^{(M)}_{\mu}(q,p,z)\,=\,0\,,\,\,\,\,\forall\,\,\,M \geq 2\,.
\ee

\noindent
Hence, Eq.(\ref{mlett:aII-15}) reduces to

\be
\label{II-23}
{\mathcal T}_{\mu}(q,p,z)\,=\,{\mathcal T}^{(0)}_{\mu}(q,p)\,+\,{\mathcal T}^{(1)}_{\mu}(q,p)\,=\,p_\mu-\Theta_{\mu\nu}q^\nu\,+\,\Delta_{\mu\nu}\,z^\nu\,,
\ee

\noindent
where we have substituted ${\mathcal T}^{(0)}_{\mu}(q,p)$ in accordance with Eq.(\ref{II-4}). Moreover, Eqs.(\ref{II-3}) and (\ref{II-91}) allow for splitting Eq.(\ref{II-23}) as follows

\bml
\label{II-24}
\bea
&&{\mathcal T}_{j}(q,p,z)\,=\,p_j\,-\,a v_j\,-\,a\,g_{jk}\,z^{N + k}\,,\label{mlett:aII-24}\\
&&{\mathcal T}_{N + j}(q,p,z)\,=\,\pi_j\,+\,a z_j\,-\,a^2\,\theta_{jk}\,\left(v^k\,+\,2 z^{N + k}\right)\,.\label{mlett:bII-24}
\eea
\eml

\noindent
The problem of finding the BFT extension of the constraints is over.

What remains to be done is to find an extension for the Hamiltonian. The fact that only ${\mathcal T}^{(1)}_{\mu}(q,p,z)$ is nonvanishing simplifies the set of recurrence relations arising from Eq.(\ref{mlett:bII-14}). For a generic $M$ one finds

\be
\label{II-25}
[ {\mathcal T}^{(0)}_\mu(q, p)\,,\,H^{(M)}(q, p, z)]_{PB_{\Sigma}}\,+\,[ {\mathcal T}_{\mu}^{(1)}(q, p, z)\,,\, H^{(M+1)}(q, p, z)]_{PB_{\Lambda}}\,=\,0\,.
\ee

\noindent
We claim that (see Eq.(\ref{II-17}))

\be
\label{II-26}
Y^{(M)}_{\alpha_1...\alpha_M}(q , p)\,=\,0\,,
\ee

\noindent
when any of the subscripts takes values in the interval $[N+1 , 2N]$, together with

\be
\label{II-27}
Y^{(M)}_{i_1...i_M}(q , p)\,=\,\frac{1}{M!} \frac{\partial^{M} H^{(0)}(q , p)}{\partial q^{i_1} \cdots \partial q^{i_M}}\,,
\ee

\noindent
solve Eq.(\ref{II-25}). If true, this implies that

\bea
\label{II-28}
{\mathcal H}(q,p,z)\, = \,H^{(0)}(q,p)\,+\,\sum_{M = 1}^{+\infty}\,H^{(M)}(q , p, z)\,\equiv :\,H^{(0)}(q + z , p)\,
\eea

\noindent
since

\be
\label{II-281}
H^{(M)}(q , p, z)\,=\,Y^{(M)}_{i_1...i_M}(q , p)\,z^{i_1}...z^{i_M}\,=\,\frac{1}{M!} \frac{\partial^{M} H^{(0)}(q , p)}{\partial q^{i_1} \cdots \partial q^{i_M}}\,z^{i_1}...z^{i_M}\,,
\ee

\noindent
as it follows from Eqs.(\ref{mlett:bII-15}), (\ref{II-17}), (\ref{II-26}) and (\ref{II-27}). Thus, the extension of the Hamiltonian is obtained by translating $ q^i \longrightarrow q^i + z^i$. It remains to be shown that the assertions made in this paragraph indeed hold. To that end we shall follow a two steps procedure. We first set $\mu = i$ in Eq.(\ref{II-25}) which, then, goes into

\be
\label{II-29}
[ {\mathcal T}^{(0)}_i(q , p)\,,\,H^{(M)}(q, p, z)]_{PB_{\Sigma}}\,+\,[ {\mathcal T}_i^{(1)}(q, p, z)\,,\, H^{(M+1)}(q, p, z)]_{PB_{\Lambda}}\,=\,0\,.
\ee

\noindent
It must be kept in mind that ${\mathcal T}^{(0)}_i(q , p)$ as well as $H^{(0)}(q , p)$ only depend on those $q^i$'s and $p_i$'s for which $i \leq N$. In view of Eq.(\ref{II-281}), the same applies for $H^{(M)}(q , p, z)$. Let us consider the problem of evaluating the first term in the left hand side of Eq.(\ref{II-29}). By bringing into it the explicit form in Eq.(\ref{II-4}) and after recalling Eqs.(\ref{II-3}) one arrives at

\be
\label{II-30}
[ {\mathcal T}^{(0)}_i(q , p)\,,\,H^{(M)}(q, p, z)]_{PB_{\Sigma}}\,=\,-\,\frac{\partial H^{(M)}(q, p, z)}{\partial q^i}\,.
\ee

\noindent
The evaluation of the second term in the left hand side of Eq.(\ref{II-29}) is more involved. To begin with one reads ${\mathcal T}_i^{(1)}(q, p, z)$ directly from Eq.(\ref{II-23}) and, therefore, writes

\bea
\label{II-31}
\left[ {\mathcal T}_i^{(1)}(q, p, z)\,,\,H^{(M + 1)}(q , p, z)\right]_{PB_{\Lambda}}\,=\,Y^{(M + 1)}_{i_1...i_{M + 1}}(q , p)\,\Delta_{i \nu}\,  [z^{\nu}\,,\,z^{i_1}...z^{i_{M + 1}}]\,,
\eea

\noindent
where Eq.(\ref{II-281}) has been taken into account. The computation of the commutator in the right hand side of this last equation, which requires the repeated use of Eq.(\ref{II-12}), yields

\bea
\label{II-32}
\Delta_{i \nu}\,[z^{\nu}\,,\,z^{i_2}...z^{i_{M + 1}}]_{PB_{\Lambda}}\,=\, \delta_i^{\,\,i_1} z^{i_2} \cdots z^{i_{M + 1}}\,+\, \ldots \,+\,z^{i_1}...z^{i_{M}}\,\delta_i^{\,\,i_{M + 1}}\,.
\eea

\noindent
We observe that $Y^{(M)}_{i_1...i_M}(q , p)$ is symmetric under the exchange of any pair of indices. This greatly simplifies the expression arising after the replacement of Eq.(\ref{II-32}) into Eq.(\ref{II-31}). It is found to read

\bea
\label{II-33}
&&\left[ {\mathcal T}_i^{(1)}(q, p, z)\,,\,H^{(M + 1)}(q , p, z)\right]_{PB_{\Lambda}}\,=\,(M + 1) Y^{(M + 1)}_{i i_1...i_M}(q , p)\, z^{i_1}...z^{i_{M}}\nonumber\\
&&=\,\frac{\partial Y^{(M)}_{i_1...i_M}(q , p)}{\partial q^i} z^{i_1}...z^{i_{M}}\,=\,+\,\frac{\partial H^{(M)}(q, p, z)}{\partial q^i}\,.
\eea

\noindent
Clearly, Eqs.(\ref{II-30}) and (\ref{II-33}) assert the validity of Eq.(\ref{II-29}). Secondly, we set $\mu = N + i$ in Eq.(\ref{II-25}) and, thus, obtain

\be
\label{II-34}
[ {\mathcal T}^{(0)}_{N + i}(q , p)\,,\,H^{(M)}(q, p, z)]_{PB_{\Sigma}}\,+\,[ {\mathcal T}_{N + i}^{(1)}(q, p, z)\,,\, H^{(M+1)}(q, p, z)]_{PB_{\Lambda}}\,=\,0\,.
\ee

\noindent
Furthermore, Eqs.(\ref{II-3}) and (\ref{II-4}) lead to

\bea
\label{II-35}
[ {\mathcal T}^{(0)}_{N + i}(q , p)\,,\,H^{(M)}(q, p, z)]_{PB_{\Sigma}}\,=\,[\pi_i - a^2 \theta_{ik} v^k \,,\,H^{(M)}(q, p, z)]_{PB_{\Sigma}}\,=\,0\,,
\eea

\noindent
since $H^{(M)}(q, p, z)$ does not depend on the variables belonging to the sector $N+1 \leq \mu \leq 2N$. As for the evaluation of the second term in the left hand side of Eq.(\ref{II-34}), we invoke Eqs.(\ref{II-23}), (\ref{II-281}) and (\ref{II-12}) to get

\bea
\label{II-36}
&&[ {\mathcal T}_{N + i}^{(1)}(q, p, z)\,,\, H^{(M+1)}(q, p, z)]_{PB_{\Lambda}}\,=\,Y^{M + 1}_{i_1 \cdots i_{M + 1}}(q , p)\,\Delta_{N + i\,\, \nu}\,\left[z^{\nu}\,,\,z^{i_1} \ldots z^{i_{`M + 1}} \right]\nonumber\\
&&=\,Y^{M + 1}_{i_1 \ldots i_{M + 1}}(q , p)\,\left( \,\delta_{N + i}^{\,\,i_1} z^{i_2} \cdots z^{i_{M + 1}}\,+\, \cdots \,+\,z^{i_1}...z^{i_{M}}\,\delta_{N + i}^{\,\,i_{M + 1}}\right)\,=0\,,
\eea

\noindent
since the indices in each Kronecker symbol belong to non overlapping sectors. Therefore, the left hand side of Eq.(\ref{II-34}) vanishes identically which completes the purported proof.

To summarize, we have presented in this Section the BFT embedding of the second class theory that gives origin to noncommutative quantum mechanics. Our findings are not unique but, however, a different choice for $\omega^{\mu \nu}$ leads to an extension differing from ours by a canonical transformation \cite{Bat,Frad,Tyu}.

\section{The gauge invariant sector}
\label{sec:level3}

Let us see what we learn from the counting of the degrees of freedom in the second class model as well as in its BFT extension. The number of dimensions of the phase space of the second class theory is $d[\Sigma] = 4 N$ while the number of independent phase space variables is $4N - 2N = 2N$, being $2N$ the number of second class constraints. As for the gauge theory, the number of dimensions of its phase space is $d[\Gamma] = 8 N$ whereas the number of independent phase space variables is $8N - 4N = 4N$, where $4N$ includes the first class constraints and the gauge conditions. Therefore, it is not self-evident that both models describe the same physical reality. To show that this is indeed the case we shall start by constructing the physical phase space which is the one spanned by gauge independent degrees of freedom. We shall derive, afterwards, the $PB$ algebra fulfilled by these phase space coordinates. Also, the constraints and the Hamiltonian will be written in terms of gauge invariant phase space variables. As we shall see, through this procedure one uniquely recovers the Hamiltonian formulation of the second class model.

To begin with, we recall that the generator of infinitesimal gauge transformations ($G$) is given by the expression \cite{Fradkin1,Girotti3}

\be
\label{III-1}
G\,=\,\e^{\mu} {\mathcal T}_{\mu}\,,
\ee

\noindent
where $\e^{\mu}, \mu = 1, \ldots , 2N$, denote a set of independent infinitesimal gauge parameters and ${\mathcal T}_{\mu}$ is given at Eq.(\ref{II-23}). Then, under infinitesimal gauge transformations the $q$'s, $p$'s and $z$'s change, respectively, as

\bml
\label{III-2}
\bea
&&{\bar \delta}q^{\mu}\,=\,[q^{\mu} , G]_{PB_{\Sigma}}\,=\,\e^{\mu}\,,\label{mlett:aIII-2}\\
&&{\bar \delta}p_{\mu}\,=\,[p_{\mu} , G]_{PB_{\Sigma}}\,=\,\Theta_{\rho \mu}\e^{\rho} \Longrightarrow \left\{ \begin{array}{l}
{\bar \delta}p_{i}\,=\,0\\
{\bar \delta}\pi_{i}\,=\,a \e_i - a^2 \theta_{ik} \e^{N + k}
\end{array} \right. \,,\label{mlett:bIII-2}\\
&&{\bar \delta}z^{\mu}\,=\,[z^{\mu} , G]_{PB_{\Lambda}}\,=\,-\,\e^{\mu}\,.\label{mlett:cIII-2}
\eea
\eml

\noindent
Hence, the composite objects $Q^{\mu}$ and $P_{\mu}$,

\bml
\label{III-3}
\bea
&&Q^{\mu}\,\equiv\,q^{\mu}\,+\,z^{\mu}\,,\label{mlett:aIII-3}\\
&&P_{\mu}\,\equiv\,p_{\mu}\,+\,z^{\nu} \Theta_{\nu \mu} \Longrightarrow \left\{ \begin{array}{l}
P_i\,=\,p_i\\
P_{N + i}\,=\,\pi_i + a z_i\,-\,a^2 \theta_{ik}\,z^{N + k}
\end{array}
\right.\label{mlett:bIII-3}\,,
\eea
\eml

\noindent
remain invariant under gauge transformations. We conjecture that they serve for spanning the physical phase space. To confirm this we first evaluate their $PB$ algebra. The calculations are straightforward and yield

\bml
\label{III-4}
\bea
&&[Q^{\mu}\,,\,Q^{\nu}]_{PB_{\Gamma}}\,=\,\left(\Delta^{-1} \right)^{\mu \nu}\,,\label{mlett:aIII-4}\\
&&[Q^{\mu}\,,\,P_{\nu}]_{PB_{\Gamma}}\,=\,\delta^{\mu}_{\,\,\nu}\,+\, \left(\Delta^{-1} \right)^{\mu \alpha} \Theta_{\alpha \nu}\,,\label{mlett:bIII-4}\\
&&[P_{\mu}\,,\,P_{\nu}]_{PB_{\Gamma}}\,=\,\Theta_{\alpha \mu}\left(\Delta^{-1} \right)^{\alpha \beta} \Theta_{\beta \nu}\,=\,0\,,\label{mlett:cIII-4}
\eea
\eml

\noindent
which exactly duplicates the $DB$ algebra of the corresponding variables spanning the phase space of the second class system (see Eqs.(\ref{II-7})-(\ref{II-9})). We emphasize that the comparison is between $PB$'s involving gauge invariant quantities belonging to the first class system with $DB$'s involving the corresponding counterparts in the second class one.

However, to establish the equivalence between the original second class theory and its BFT extension demands further work. In fact, we must investigate the form assumed by the first class constraints and the Hamiltonian when written in terms of the gauge invariant phase space coordinates defined at Eq.(\ref{III-3}). Let us first concentrate on the constraints. We notice that Eqs.(\ref{II-23}), (\ref{II-5}) and (\ref{III-3}) lead to

\be
\label{III-5}
{\mathcal T}_{\mu}(q , p, z)\,=\,P_{\mu}\,-\,\Theta_{\mu \nu}\,Q^{\nu}\,=\,{\mathcal T}_{\mu}^{(0)}(Q , P)\,,
\ee

\noindent
in agreement with Eq.(\ref{II-4}). As for the Hamiltonian, Eqs.(\ref{II-28}) and (\ref{III-3}) yield

\be
\label{III-6}
{\mathcal H}(q, p, z)\,=\,H^{(0)}(Q , P)\,,
\ee

\noindent
which completes the desired proof of equivalence. In fact, we have carried out a complete reconstruction of the Hamiltonian formulation of the dynamics of the initial second class model.

One should notice that the transition from $q$, $p$, $z$ to $Q$ and $P$ implies in a dimensional reduction process and, hence, it is not a canonical transformation.

\section{Functional quantization and quantum equivalence}
\label{sec:level4}

We shall next be concerned with the equivalence between the quantized version of the second class model and that associated with the gauge theory arising from the BFT embedding.

It follows from the specialized literature on systems with constraints \cite{Fradkin1,Girotti2,Sundermeyer1,Gitman1} that the phase space path integral yielding the generating functional of Green functions (${\mathcal Z}$) is, in the case of the second class model,

\bea
\label{IV-1}
{\mathcal Z}\,&=&\,{\mathcal N}\,\int \left[{\mathcal D}^{2N}q \right] \int \left[{\mathcal D}^{2N}p \right] \,\left(\prod_{\mu = 1}^{2N} \delta \left[ {\mathcal T}_{\mu}^{(0)}(q , p)\right] \right)\,\left(\prod_t \det \|\Delta \| \right)^{\frac {1}{2}}\nonumber\\
&\times&\,\exp\left\{ \frac{i}{\hbar} \int_{t_{in}}^{t_f} dt\,\left[p_{\mu} {\dot q}^{\mu}\,-\,H^{(0)}(q , p) \right]\right\}\nonumber\\
&=&{\mathcal N} \int \left[{\mathcal D}^{2N}q \right] \int \left[{\mathcal D}^{2N}p \right] \left(\prod_{\mu = 1}^{2N} \delta \left[ {\mathcal T}_{\mu}^{(0)}(q , p)\right] \right) \nonumber\\
&\times& \exp\left\{ \frac{i}{\hbar} \int_{t_{in}}^{t_f} dt \left[p_{\mu} {\dot q}^{\mu} - H^{(0)}(q^i , p_i) \right]\right\}\nonumber\\
&=& {\mathcal N} \int \left[{\mathcal D}^{N}q \right] \int \left[{\mathcal D}^{N}p \right] \exp\left\{ \frac{i}{\hbar} \int_{t_{in}}^{t_f} dt \left[p_i {\dot q}^{i} + {\dot p}_i \theta^{ij} p_j - H^{(0)}(q^i , p_i) \right]\right\}\,,
\eea

\noindent
where in going from the second to the third term of the equality we took into account that $\|\Delta\|$ is an irrelevant constant matrix (see Eq.(\ref{II-6})). Moreover, for arriving to the last term of the equality we explore the fact that the lack of dependence of $H^{(0)}$ upon the variables $q^{\mu}, p_{\mu}, \mu = N + 1, \ldots,2N$, allows for using the constraints to integrate out the just mentioned phase space sector.

As for the gauge theory obtained through the BFT scheme one has that \cite{Fradkin1,Girotti2,Sundermeyer1,Gitman1}

\bea
\label{IV-2}
{\mathcal Z}_{\chi}\,&=&\,{\mathcal N}_{\chi}\,\int \left[{\mathcal D}^{2N}q \right] \int \left[{\mathcal D}^{2N}p \right] \int \left[{\mathcal D}^{2N}u \right] \int \left[{\mathcal D}^{2N}s \right]\nonumber\\
&\times& \left(\prod_{\mu = 1}^{2N} \delta \left[ {\mathcal T}_{\mu}(q , p, z)\right] \right) \left(\prod_{\mu = 1}^{2N} \delta \left[ \chi^{\mu}(q , p, z)\right] \right) \left(\prod_t \det\left[{\mathcal T}_{\mu} , \chi^{\nu} \right]_{PB_{\Gamma}}  \right)\nonumber\\
&\times&  \exp\left\{ \frac{i}{\hbar} \int_{t_{in}}^{t_f} dt \left[p_{\mu} {\dot q}^{\mu} + s_{\mu} {\dot u}^{\mu} - H^{(0)}(q^i + z^i , p_i) \right]\right\}\,.
\eea

\noindent
Here, ${\mathcal N}_{\chi}$ is a normalization constant whereas $\chi = \chi(q , p, z)$ denotes a set of arbitrarily chosen gauge functions. The question now posses itself: does the right hand side of Eq.(\ref{IV-2}) falls back into that in Eq.(\ref{IV-1}), for any $\chi$? We shall illustrate the situation for two different gauges.

The first gauge to be analyzed, commonly referred to as the unitary gauge, is specified by the subsidiary conditions

\be
\label{IV-4}
\chi^{\nu}\,=\,z^{\nu}\,\approx\,0\,.
\ee

\noindent
Then, Eqs.(\ref{II-23}), (\ref{IV-4}), and (\ref{II-12}) lead to

\bea
\label{IV-5}
\det \left[{\mathcal T} _{\mu}\,,\,\chi^{\nu} \right]_{PB_{\Gamma}}\,=\,\delta_{\mu}^{\,\,\nu} \Longrightarrow  \left(\prod_t \det\left[T_{\mu} , \chi^{\nu} \right]_{PB_{\Gamma}}  \right)\,=\,1\,.
\eea

\noindent
We notice that the ${\mathcal T}$'s, $\chi$'s and $H$ depend on the variables $u$ and $s$ only through the combination $z^{\mu} = -1/2 u^{\mu} + (\Delta^{-1})^{\mu \nu} s_{\nu}$. This strongly suggests the convenience in performing the change of dummy integration variables $u^{\mu} \rightarrow u^{\prime \mu} = z^{\mu}$, $s_{\mu} \rightarrow s^{\prime}_{\mu} = s_{\mu}$. The path integrals on $s^{\prime}$ decouple from the rest and can be explicitly evaluated. One ends up with

\bea
\label{IV-6}
{\mathcal Z}_{z = 0}\,&=&\,{\mathcal N}\,\int \left[{\mathcal D}^{2N}q \right] \int \left[{\mathcal D}^{2N}p \right] \int \left[{\mathcal D}^{2N}z \right]
\left(\prod_{\mu = 1}^{2N} \delta \left[ {\mathcal T}_{\mu}(q , p, z)\right] \right) \left(\prod_{\mu = 1}^{2N} \delta \left[ z\right] \right) \nonumber\\
&\times&  \exp\left\{ \frac{i}{\hbar} \int_{t_{in}}^{t_f} dt \left[p_{\mu} {\dot q}^{\mu} + \frac{1}{2}z^{\mu} \Delta_{\mu \nu} {\dot z}^{\nu} - H^{(0)}(q^i + z^i , p_i) \right]\right\}\,.
\eea

\noindent
The $z$-integration is straightforward and yields 

\bea
\label{IV-7}
{\mathcal Z}_{z = 0}\,&=&\,{\mathcal N}\,\int \left[{\mathcal D}^{2N}q \right] \int \left[{\mathcal D}^{2N}p \right] \left(\prod_{\mu = 1}^{2N} \delta \left[ {\mathcal T}^{(0)}_{\mu}(q , p)\right] \right)\nonumber\\
&&\exp\left\{ \frac{i}{\hbar} \int_{t_{in}}^{t_f} dt \left[p_{\mu} {\dot q}^{\mu}\,-\,H^{(0)}(q^i , p_i) \right]\right\}\,.
\eea

\noindent
As in the case of the second class system we use the constraints to integrate out the variables $q^{\mu},\; p_{\mu},\; \mu = N + 1, \ldots,2N$, which send us back to Eq.(\ref{IV-1}), i.e.,

\bea
\label{IV-8}
{\mathcal Z}_{z = 0}\,&=&\,{\mathcal N}\,\int \left[{\mathcal D}^{N}q \right] \int \left[{\mathcal D}^{N}p \right] \exp\left\{ \frac{i}{\hbar} \int_{t_{in}}^{t_f} dt \left[p_i {\dot q}^{i} + {\dot p}_i \theta^{ij} p_j  - H^{(0)}(q^i , p_i) \right]\right\}\,.
\eea

For our present purposes, the unitary gauge is the easiest one to deal with because the gauge conditions (\ref{IV-4}) explicitly kill the variables responsible for bringing the gauge freedom into play. On the other hand, it is also worth mentioning that our proof of equivalence is model independent. This is due to the fact that all that is required for obtaining the BFT extension of the Hamiltonian is to perform the translation $q \longrightarrow q + z$.

Next in the sequel is the gauge

\be
\label{IV-9}
\chi^{\nu}\,=\,q^{\nu}\,\approx\,0\,.
\ee

\noindent
By starting from Eqs.(\ref{II-23}) and (\ref{IV-9}) we obtain again a functional determinant that reduces to a nonvanishing irrelevant constant, i.e.,

\bea
\label{IV-10}
\det \left[{\mathcal T} _{\mu}\,,\,\chi^{\nu} \right]_{PB_{\Gamma}}\,=\,-\,\delta_{\mu}^{\,\,\nu} \Longrightarrow  \left(\prod_t \det\left[T_{\mu} , \chi^{\nu} \right]_{PB_{\Gamma}}  \right)\,=\,\mbox{nonvanishing constant}\,.
\eea

\noindent
As for the results obtained in connection with the integration on the variables $s$ they remain as before. This enables us to write Eq.(\ref{IV-2}) as

\bea
\label{IV-11}
{\mathcal Z}_{q = 0}\,&=&\,{\mathcal N}\,\int \left[{\mathcal D}^{2N}q \right] \int \left[{\mathcal D}^{2N}p \right] \int \left[{\mathcal D}^{2N}z \right] \nonumber\\
&\times& \left(\prod_{\mu = 1}^{2N} \delta \left[ p_\mu-\Theta_{\mu\nu}q^\nu\,+\,\Delta_{\mu\nu}\,z^\nu\right] \right) \left(\prod_{\mu = 1}^{2N} \delta \left[ q \right] \right) \nonumber\\
&\times&  \exp\left\{ \frac{i}{\hbar} \int_{t_{in}}^{t_f} dt \left[p_{\mu} {\dot q}^{\mu} + \frac{1}{2}z^{\mu} \Delta_{\mu \nu} {\dot z}^{\nu}  - H^{(0)}(q^i + z^i , p_i) \right]\right\}\,,
\eea

\noindent
where Eq.(\ref{II-23}) has been taken into account. The integrations on $q$'s and $p$'s can be carried out at once and yield

\bea
\label{IV-12}
{\mathcal Z}_{q = 0}\,&=&\,{\cal N}\int \left[{\cal D}^{2N} z\right]\,
\exp\left\{\frac{i}{\hbar}\int_{t_{in}}^{t_f} dt\left[\frac{1}{2} z^\mu\Delta_{\mu\nu}\dot{z}^\nu\,-\,H^{(0)}(z^i,a z_{N +i})\right]\, \right\}\nonumber\\
&=&{\cal N}\int \left[ {\cal D}^{2N} z \right]\,
\exp \left\{\frac{i}{\hbar}\int_{t_{in}}^{t_f} dt\left[\frac{a}{2}\,z_{N + i} {\dot z}^i\, -\, \frac{a}{2}\, z^i {\dot z}_{N + i}\right.\right.\nonumber\\
& +&\left. \left.a^2 {\dot z}_{N + i} \theta^{ij} z_{N + j}\, - \,H^{(0)}(z^i,a z_{N +i})\right] \right\}
\eea

\noindent
which, after relabeling $ a z_{N + i} \rightarrow p_i$, $z^i \rightarrow q^i$ and neglecting a surface term, goes into

\bea
\label{IV-13}
{\mathcal Z}_{q = 0}\,&=&\,{\cal N}\int \left[{\cal D}^{N} q\right]\,\int \left[{\cal D}^{N} p\right]\,\exp\left\{\frac{i}{\hbar}\int_{t_{in}}^{t_f} dt\left[p_i {\dot q}^i + {\dot p}_i \theta^{ij} p_j \,-\,H^{(0)}(q^i, p_i\right] \right\}
\eea

\noindent
in agreement with Eq.(\ref{IV-1}). Again, the proof of equivalence does not call for restrictions on the structure of the Hamiltonian.

Up to this point we have, loosely speaking, analyzed the quantum equivalence between the second and the first class theories for two kind of {\it extreme} gauges. Firstly, the gauge conditions were chosen so as to eliminate the variables that were not present in the second class model. It is then natural to expect that the gauge theory falls back into the second class one. Secondly, the gauge conditions set to zero the basic variables ($q$). The corresponding canonically conjugate momenta ($p$) were also integrated out. Nevertheless, the variables $z^i$ and $z_{N + i}$ took over, respectively, the roles of $q$ and $p$ allowing for the reconstruction of the original second class theory.

\section{Operator quantization}
\label{sec:level5}

What comes next is the quantization of the gauge theory within the operator approach and its relationship with the outcomes obtained for the second class model when subjected to the same scheme of quantization.

The gauge theory will be quantized by using the method put forward by Dirac \cite{Dirac1} which, unlike the functional method, does not demand for the elimination of the gauge freedom. The main ingredients are: i) the physical states ($\Ket{\Psi(t)}$) are required to fulfil

\be
\label{V-6}
\hat{\mathcal T}_\mu(\hat{Q}^\mu,\hat{P}_\nu,\hat{Z}^\lambda)\Ket{\Psi}=0\,,
\ee

\noindent
while ii) the dynamics is controlled by the equation

\be
\label{V-7}
\hat{H}^{(0)}(\hat{Q}^i+\hat{Z}^i, \hat{P}_i)\Ket{\Psi(t)}\,=\,i\hbar\frac{d\Ket{\Psi(t)}}{dt}\,.
\ee

\noindent
Needless to say, $\hat{\mathcal T}_\mu(\hat{Q}^\mu,\hat{P}_\nu,\hat{Z}^\lambda)$ and $\hat{H}^{(0)}(\hat{Q}^i+\hat{Z}^i, \hat{P}_i)$ are the quantum counterparts of Eqs.~(\ref{II-23}) and (\ref{II-28}), respectively. We recall that, within Dirac's method of quantization, the basic operators obey an equal time algebra abstracted from the corresponding Poisson bracket algebra, i.e.,

\be
\label{V-71}
[\hat{A},\hat{B}]\,=\,i\,\hbar\,[a,b]_{PB_{\Gamma}}\Bigr|_{\begin{array}{l} a \rightarrow {\hat A}\\ b \rightarrow {\hat B} \end{array}}\,,
\ee

\noindent
where $a$ and $b$ may either be $q^i$, $v^i$, $p_i$, $\pi_i$, $u^\mu$ or $s_\mu$, while $\hat{A}$ and $\hat{B}$ may either denote $\hat{Q}^i$, $\hat{V}^i$, $\hat{P}_i$, $\hat{\Pi}_i$, $\hat{U}^\mu$ or $\hat{S}_\mu$.

It is not difficult to convince oneself that the above implies that Eqs.~(\ref{II-28}) and (\ref{II-281}) may be cast, at the quantum level,

\be
\label{V-8}
\hat{H}^{(0)}(\hat{Q}^i+\hat{Z}^i, \hat{P}_i)
\,=\,\sum_{M=0}^\infty\frac{1}{M!}\frac{\partial^M \hat{H}^{(0)}(\hat{Q}^i, \hat{P}_i)}{\partial
\hat{Q}^{i_1}\cdots\partial\hat{Q}^{i_M}}\hat{Z}^{i_1}\cdots\hat{Z}^{i_M}\,.
\ee

\noindent
We emphasize that any reordering in the right rand side of Eq.~(\ref{V-8}) would give rise to the appearance of products of c-number antisymmetric factors $i\hbar(\Delta^{(-1)})^{ij}$ (see Eq.(\ref{II-12})) which can be disregarded in view of the symmetry of the coefficient operator $\frac{\partial^M \hat{H}^{(0)}(\hat{Q}^i, \hat{P}_i)}{\partial
\hat{Q}^{i_1}\cdots\partial\hat{Q}^{i_M}}$.

Let us now return to Eq.(\ref{V-6}). After solving for $z^i$ from Eq.(\ref{II-23}) and then transferring the result to the quantum level one finds

\be
\label{V-9}
\hat{Z}^i\Ket{\Psi}=\left(-a\theta^{ij}\hat{V}_j+2\theta^{ij}\hat{P}_j-\frac{1}{a}\hat{\Pi}^i\right)\Ket{\Psi}\,,
\ee

\noindent
and, whence,

\bea
\label{V-10}
\hat{Z}^{i_1}\cdots \hat{Z}^{i_M}\Ket{\Psi}
&=& \left(-a\theta^{i_Mj_M}\hat{V}_{j_M}+2\theta^{i_Mj_M}\hat{P}_{j_M}-\frac{1}{a}\hat{\Pi}^{i_M}\right)\nonumber\\
&\times&\cdots\left(-a\theta^{i_1j_1}\hat{V}_{j_1}+2\theta^{i_1j_1}\hat{P}_{j_1}-\frac{1}{a}\hat{\Pi}^{i_1}\right)\Ket{\Psi}\,.
\eea

\noindent
By substituting Eq.(\ref{V-10}) into Eq.(\ref{V-8}) and the result thus obtained into Eq.(\ref{V-7}) one gets

\be
\label{V-11}
\hat{H}^{(0)}\left(\hat{Q}^i-a\theta^{ij}\hat{V}_j+2\theta^{ij}\hat{P}_j-\frac{1}{a}\hat{\Pi}^i, \hat{P}_i\right)\Ket{\Psi(t)}\,=\,i\hbar\frac{d\Ket{\Psi(t)}}{dt}\,,
\ee

\noindent
which does not longer involves the phase space variables $\hat{U}$ and $\hat{S}$. We observe that the gauge freedom has been eliminated without recourse to subsidiary conditions.

One may check that the new variables

\bml
\label{V-12}
\bea
&&{\hat {Q}}^{\prime\,i}\,\equiv\,\hat{Q}^i-a\theta^{ij}\hat{V}_j+2\theta^{ij}\hat{P}_j-\frac{1}{a}\hat{\Pi}^i\,,
\label{mlett:aV-12}\\
&&\hat{P}^{\prime}_i\,\equiv\,\hat{P}_i\,,\label{mlett:bV-12}
\eea
\eml

\noindent
obey the equal-time commutator algebra in Eq.(\ref{I-1}). Furthermore, in terms of them Eq.(\ref{V-11}) acquires the form

\be
\label{V-13}
\hat{H}^{(0)}\left({\hat {Q}}^{\prime\,i},\hat{P}^{\prime}_i \right)\Ket{\Psi(t)}\,=\,i\hbar\frac{d\Ket{\Psi(t)}}{dt}\,,
\ee

\noindent
which reproduces the time evolution of the quantized second class system. Again, the equivalence becomes established.

\section{Conclusions}
\label{sec:level6}

This work was dedicated to formulate noncommutative quantum mechanics as a gauge theory.  The tool for carrying out that task was the BFT embedding procedure. The extensions of the constraints and of the Hamiltonian gives origin to an involution algebra defining an Abelian gauge theory. We provided a detailed and rigorous proof of the consistency of this formulation by demonstrating that the initial second class system can be uniquely recovered from the gauge invariant sector of the gauge theory. This confirms the equivalence of the second and first class formulations at the classical level.

The quantization of the gauge extension within the functional framework follows along the standard lines. The flexibility offered by the gauge choice opens new avenues for carrying out explicit model calculations. We carried out the quantization in two specific gauges. For both of them it was possible to show that the phase space functional integral yielding the Green functions generating functional maps into the corresponding one arising in connection with the second class model. Moreover, the equivalence between  the alternative descriptions of noncommutative quantum mechanics turns out to be model independent.

The quantization of the first class model within the operator approach was implemented by using the formalism put forward by Dirac \cite{Dirac1}. Its outcome naturally convey to the formulation of the quantum dynamics of the second class system within this scheme of quantization.

\vspace{1cm}

{\bf Acknowledgements.} We are gratefully indebted to Professors M. Gomes, A. J. da Silva and V. O. Rivelles, and to Dr. V. G. Kupriyanov for very useful discussions. We are also thankful to the Departamento de Física-Matemática do Instituto de Física da Universidade de São Paulo, where this paper was partially written, for the warm hospitality extended to us.  Both authors acknowledge partial support from Conselho Nacional de Desenvolvimento Científico e Tecnológico (CNPq), Brazil.


\newpage

\begin{thebibliography}{99}

\bibitem{Chaichian1} Chaichian M, Sheikh-Jabbari M and Tureanu A, Eur. Phys. J. C {\bf 36}, 251 (2004).

\bibitem {Gamboa1} J. Gamboa, M. Loewe, and J. C. Rojas, Phys. Rev. D {\bf 64}, 067901 (2001).

\bibitem {Gamboa2} J. Gamboa, M. Loewe, F. Mendez, and J. C. Rojas, Int. J. Mod. Phys. A {\bf 17}, 2555 (2002).

\bibitem {Girotti1} H. O. Girotti, Am. J. Phys. {\bf 72}, 608 (2004).

\bibitem {Bemfica} F. S. Bemfica and H. O. Girotti,  J. Phys. A: Math. Gen. {\bf 38}, L539 (2005).

\bibitem {Bemfica1} F. S. Bemfica and H. O. Girotti, Phys. Rev. D{\bf 77}, 027704 (2008).

\bibitem {Bemfica2} F. S. Bemfica and H. O. Girotti, Braz. J. Phys. {\bf 38}, 227 (2008).

\bibitem {Bemfica3} F. S. Bemfica and H. O. Girotti, Phys. Rev. D{\bf 78}, 125009 (2008).

\bibitem {footnote2} Out from the vast literature existing on constrained systems we mention Refs.\cite{Dirac1,Fradkin1,Sundermeyer1,Girotti2,Gitman1,Sudarshan1}.

\bibitem {Dirac1} P. A. M. Dirac, {\it Lectures on Quantum
Mechanics}, (Belfer Graduate School of Sciences, Yeshiva University,
New York, 1964).

\bibitem {Fradkin1} E. S. Fradkin and G. A. Vilkovisky, Phys. Lett. B{\bf 55}, 224 (1975).

\bibitem {Sundermeyer1} K. Sundermeyer, {\it Constrained Dynamics}
(Springer-Verlag, Berlin, 1982).

\bibitem {Girotti2} H. O. Girotti, ``Classical and quantum dynamics
of constrained systems,'' in {\sl Lectures in Proc. of the V Jorge
Andre Swieca Summer School}, edited by O. J. P. Eboli, M. Gomes,
and A. Santoro (World Scientific, Singapore, 1990) p1-77.

\bibitem {Gitman1}  D. M. Gitman and I. V. Tyutin, {\it Quantization
of Fields with Constraints} (Springer-Verlag, 1990).

\bibitem {Sudarshan1} E. G. C. Sudarshan and N. Mukunda, {\it
Classical Mechanics: A Modern Perspective} (John Wiley \& Sons, 1974).

\bibitem {Deriglazov1} A. A. Deriglazov, ``Noncommutative version
of an arbitrary nondegenerate mechanics,'' hep-th/0208072. Also see, C. Durval and P. Horvarty, J. Phys. A {\bf 34}, 10097 (2001).

\bibitem {footnote3} The expression in Eq.(\ref{I-2}) differs from the corresponding one in Ref.\cite{Deriglazov1} by the presence of the constant $a$ which has been introduced for homogeneizing the dimensions of the different quantities entering $L$ . The dimensions of $a$ are $g s^{-1}$.

\bibitem {Bat} I. A. Batalin and E. S. Fradkin, Nucl. Phys. B{\bf279}, 514 (1987).

\bibitem {Frad} I. A. Batalin and E. S. Fradkin, Phys. Lett. B{\bf 180}, 157 (1986).

\bibitem {Tyu} I. A. Batalin and I. V. Tyutin, Int. J. Mod. Phys. A{\bf 6}, 3255 (1991).

\bibitem {Fleck} M. Fleck and H. O. Girotti, Int. J. Mod. Phys. A{\bf 14}, 4287 (1999) and references there in.

\bibitem {footnote4} The phase space spanned by the $q$'s and $p$'s will be referred to as $\Sigma$ while $\Lambda$ labels the phase space spanned by the $u$'s and $s$'s. The full phase space is, then, $\Gamma \equiv \Sigma \oplus \Lambda$.

\bibitem {footnote5} The extra subscript $\Lambda$ specifies the phase space where the $PB$ operation is being taken. We shall continue to do so hereafter.

\bibitem {Barcelos} The possibility of implementing an embedding formalism leading to a weak (non-Abelian) involution algebra was investigated in J. Barcelos Neto and W. Oliveira, Phys. Rev. D{\bf56}, 2257 (1997).

\bibitem {Girotti3} M. E. V. Costa, H. O. Girotti and T. J. M. Simões, Phys. Rev. D{\bf 32}, 405 (1985).

\bibitem {Faddeev} L. D. Faddeev, Theor. and Math. Phys. {\bf 1}, 1 (1969).

\end{thebibliography}
\end{document}